\documentclass[graybox, envcountchap]{svmult}

\usepackage{mathptmx}        
\usepackage{amsmath}
\usepackage{amssymb}
\usepackage{color}
\usepackage{helvet}          
\usepackage{courier}         
\usepackage{dirtree}

\usepackage{makeidx}        
\usepackage{graphicx}        
\usepackage{subfig}
\usepackage{comment}

\usepackage{multicol}        
\usepackage[bottom]{footmisc}

\usepackage{hyperref}        
\hypersetup{colorlinks=true,urlcolor=blue}

\usepackage[misc]{ifsym}
\usepackage{natbib}
\usepackage{aas_macros}

\makeindex             

\begin{document}


\title{Application of Machine Learning to 21 cm Cosmology}
\author{Hayato Shimabukuro}
\institute{Hayato Shimabukuro (\Letter) \at South-Western Institute for Astronomy Research, Key Laboratory of Survey Science of Yunnan Province, Yunnan University, Kunming, Yunnan 650500, People's Republic of China \email{shimabukuro@ynu.edu.cn}}
%
%
\maketitle

This chapter reviews applications of machine learning (ML) to redshifted 21 cm cosmology, focusing on cosmic dawn, the Epoch of Reionization, and SKA-Low science. The redshifted 21 cm line directly probes diffuse neutral hydrogen, but the measured signal is not a simple astrophysical observable: density, ionization, heating, radiation backgrounds, foreground treatment, and instrumental response are coupled. The chapter first summarizes the physical ingredients needed in later sections, including the global signal, spatial fluctuations, morphology-sensitive statistics, and the 21 cm forest. It then discusses the main barriers to interpretation: bright foregrounds, radio-frequency interference, ionospheric and calibration effects, incomplete sampling, and the cost of forward modeling in high-dimensional parameter spaces. ML applications are organized by their role in the analysis chain. Observation-domain methods act on contaminated data products; theory-domain methods accelerate or compress forward modeling; and inference-domain methods connect observables to astrophysical and cosmological parameters.

\section{Introduction}

\begin{figure}[htbp]
\centering
\includegraphics[width=1.0\linewidth]{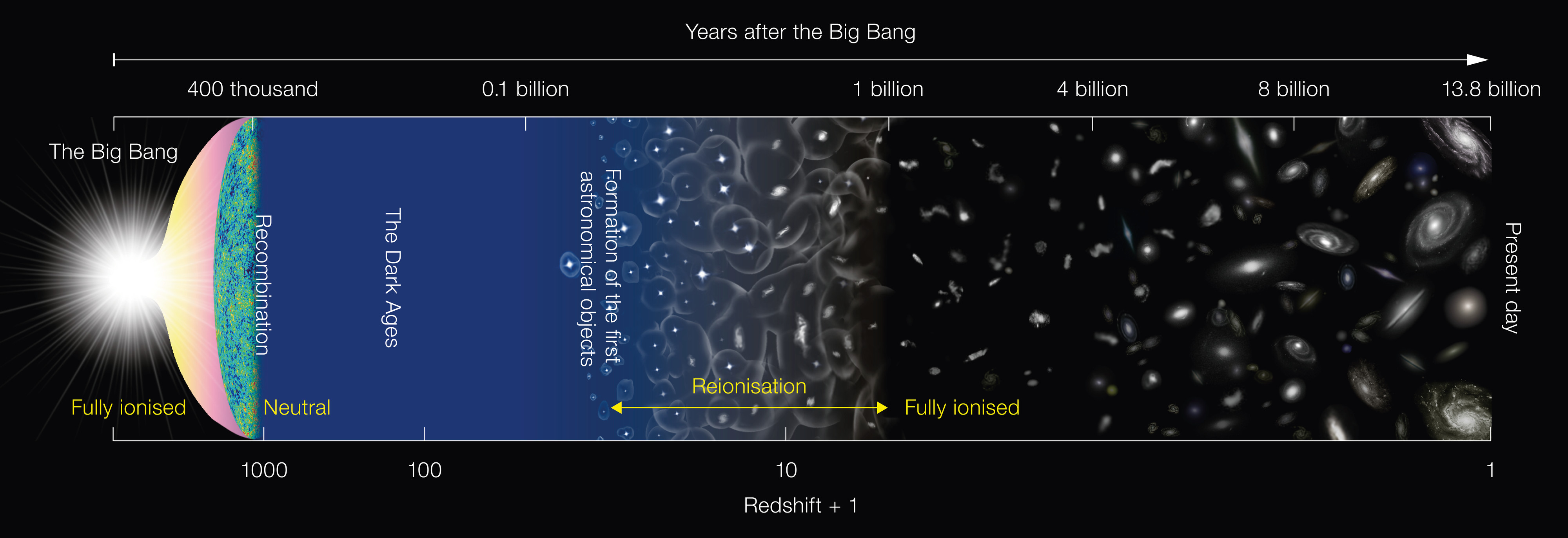}
\caption{
Schematic timeline of cosmic history from the Big Bang to the present day, highlighting the Dark Ages, the cosmic dawn, and the Epoch of Reionization.
Credit: NAOJ.
}
\label{fig:cosmic_history}
\end{figure}

Before the first stars and galaxies formed, the Universe had no luminous astrophysical sources; this period is usually called the Dark Ages. It ended gradually as the first generations of stars and galaxies began to emit radiation, heat the gas, and ionize the surrounding intergalactic medium(IGM). The cosmic dawn and the Epoch of Reionization (EoR) caused a broad transition of the IGM from a cold, mostly neutral Universe to the ionized state observed at later times. Figure~\ref{fig:cosmic_history} shows these epochs in the standard cosmic timeline. Determining when this transition occurred, and which sources drove it, remains a central goal of observational cosmology. High-redshift galaxies and quasars provide important constraints, but they do not directly map the thermal and ionization state of diffuse hydrogen over cosmological volumes.

The redshifted 21 cm line offers a direct probe of diffuse neutral hydrogen \citep{2006PhR...433..181F,2012RPPh...75h6901P,2023PASJ...75S...1S}. Global-signal experiments measure the sky-averaged brightness temperature as a function of observing frequency, whereas interferometers such as LOFAR, MWA, HERA, and SKA-Low target spatial fluctuations. The study of the 21cm signal is inherently difficult: the cosmological signal is faint, foregrounds are orders of magnitude brighter, and the relevant uncertainties enter through the instrument, the sky model, the astrophysics, and the inference procedure. Higher sensitivity in the SKA era will therefore not be sufficient by itself. Calibration, foreground control, data volume, and high-dimensional inference have to be treated as coupled parts of the same analysis chain \citep{2020PASP..132f2001L,2021MNRAS.501.1463K}.

This chapter is not a full introduction to the 21 cm line signal. We focus on the physical and statistical issues needed to assess ML applications: what information a method preserves, what approximations it introduces, and how to test those approximations.

\section{Physics of the 21 cm Signal}

We briefly explain basics of the 21cm signal below.  More detailed derivations can be found in \citep{2006PhR...433..181F,2012RPPh...75h6901P,2020PASP..132f2001L,2023PASJ...75S...1S}. Neutral hydrogen produces the 21 cm line through the hyperfine transition of its ground state. The relative abundance of the triplet and singlet states is described by the spin temperature $T_{\rm s}$,
\begin{equation}
\frac{n_{1}}{n_{0}} = 3 \exp\left( - \frac{T_{\star}}{T_{\rm s}} \right),
\end{equation}
where $n_{1}$ and $n_{0}$ are the triplet and singlet populations, and $T_{\star} = h \nu_{21} / k_{\rm B} \simeq 0.068 \ {\rm K}$ is the temperature corresponding to the hyperfine energy splitting.
If the spin temperature is equal to the cosmic microwave background (CMB) temperature, $T_{\rm s}=T_{\gamma}$, no differential 21 cm signal is observed. The signal appears when physical processes make $T_{\rm s}$ different from $T_{\gamma}$. Collisions, Lyman-$\alpha$ coupling, and X-ray heating are the most important processes during the epochs considered here:
\begin{equation}
T_{\rm s}^{-1}
=
\frac{
T_{\gamma}^{-1}
+ x_{\rm c} T_{\rm K}^{-1}
+ x_{\alpha} T_{\rm K}^{-1}
}{
1 + x_{\rm c} + x_{\alpha}
},
\end{equation}
where $x_{\rm c}$ and $x_{\alpha}$ denote the collisional and Lyman-$\alpha$ coupling coefficients, and $T_{\rm K}$ is the kinetic temperature of the gas.

A commonly used form of the differential brightness temperature is
\begin{eqnarray}
\delta T_{\rm b}(\mathbf{x}, z) &\simeq&
27 \ {\rm mK} \ 
x_{\rm HI}(\mathbf{x}, z)
\left( 1 + \delta_{\rm b}(\mathbf{x}, z) \right)
\left( 1 - \frac{T_{\gamma}(z)}{T_{\rm s}(\mathbf{x}, z)} \right)
\nonumber \\
&& \times
\left( \frac{1 + z}{10} \right)^{1/2}
\left( \frac{0.15}{\Omega_{\rm m} h^{2}} \right)^{1/2}
\left( \frac{\Omega_{\rm b} h^{2}}{0.023} \right).
\end{eqnarray}
The sign of the differential 21 cm signal is primarily governed by the ratio between the spin temperature, $T_{\rm s}$, and the background radiation temperature, $T_{\gamma}$: an absorption signature is obtained when $T_{\rm s}<T_{\gamma}$, whereas an emission signature arises when $T_{\rm s}>T_{\gamma}$. A key feature of the 21 cm observable is that it encodes, in a coupled and intrinsically nonlinear manner, the effects of the underlying matter density field, the ionization state of the intergalactic medium, the gas temperature, and the relevant radiation backgrounds \citep{2006PhR...433..181F,2012RPPh...75h6901P,2020PASP..132f2001L}. Consequently, distinct source populations and astrophysical scenarios can yield degenerate 21 cm signals that appear observationally similar. Despite these degeneracies, the 21 cm signal remains a powerful probe for extracting both cosmological and astrophysical information.

\subsection{21 cm global signal}

\begin{figure}[htbp]
\centering
\includegraphics[width=1.0\linewidth]{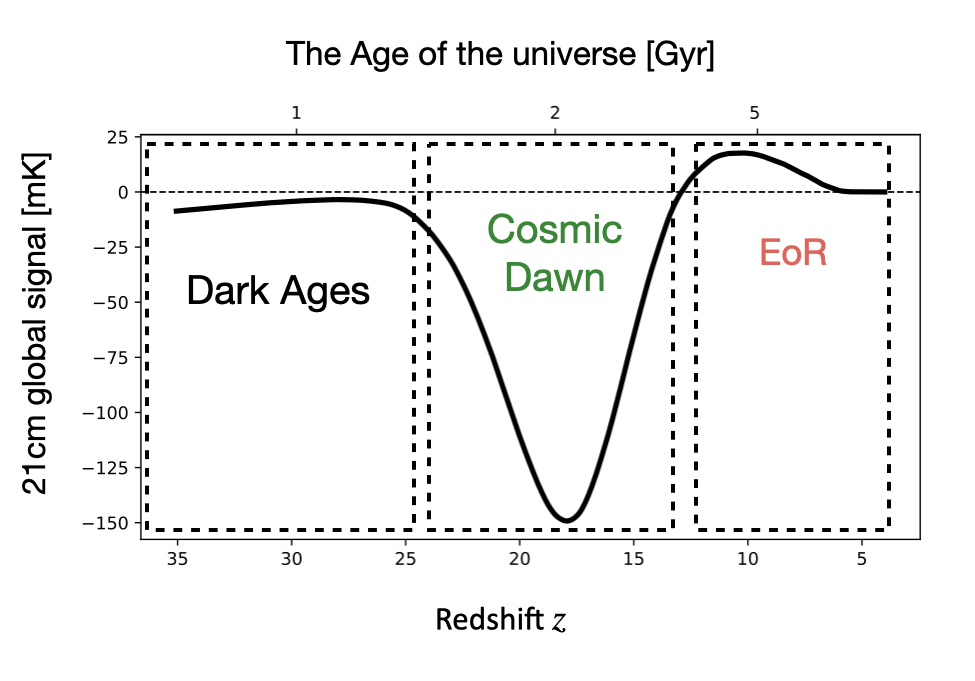}
\caption{Schematic redshift evolution of the sky-averaged 21 cm brightness temperature (the 21 cm global signal) through the Dark Ages, cosmic dawn, and the Epoch of Reionization. The schematic highlights the absorption trough during cosmic dawn, the transition to emission after X-ray heating, and the eventual disappearance of the signal as reionization progresses.}
\label{fig:global_signal}
\end{figure}

The sky-averaged 21 cm signal, usually called the global signal, is
\begin{equation}
\overline{\delta T}_{21}(z)
=
\langle \delta T_{\rm b}(\mathbf{x}, z) \rangle.
\end{equation}
The global signal gives a compact summary of the thermal and ionization history. Its evolution is often described by three broad transitions: the onset of Lyman-$\alpha$ coupling, the absorption trough before substantial X-ray heating, and the later transition toward emission once the gas is heated above the CMB temperature. Figure~\ref{fig:global_signal} shows this sequence schematically. At very high redshift, $T_{\rm s}$ remains close to $T_{\gamma}$ and the signal is weak. During cosmic dawn, radiative coupling links $T_{\rm s}$ to the colder gas temperature, producing a negative signal. X-ray heating then raises the gas temperature and can turn the signal positive. Finally, as reionization removes neutral hydrogen, the 21 cm signal fades.

The global signal constrains the integrated thermal and ionization history, but the sky average removes most spatial information. Models with different star-formation efficiencies, escape fractions, X-ray efficiencies, and source spectra can therefore produce similar global histories.

\subsection{Spatial fluctuations}

\begin{figure}[htbp]
\centering
\includegraphics[width=1.0\linewidth]{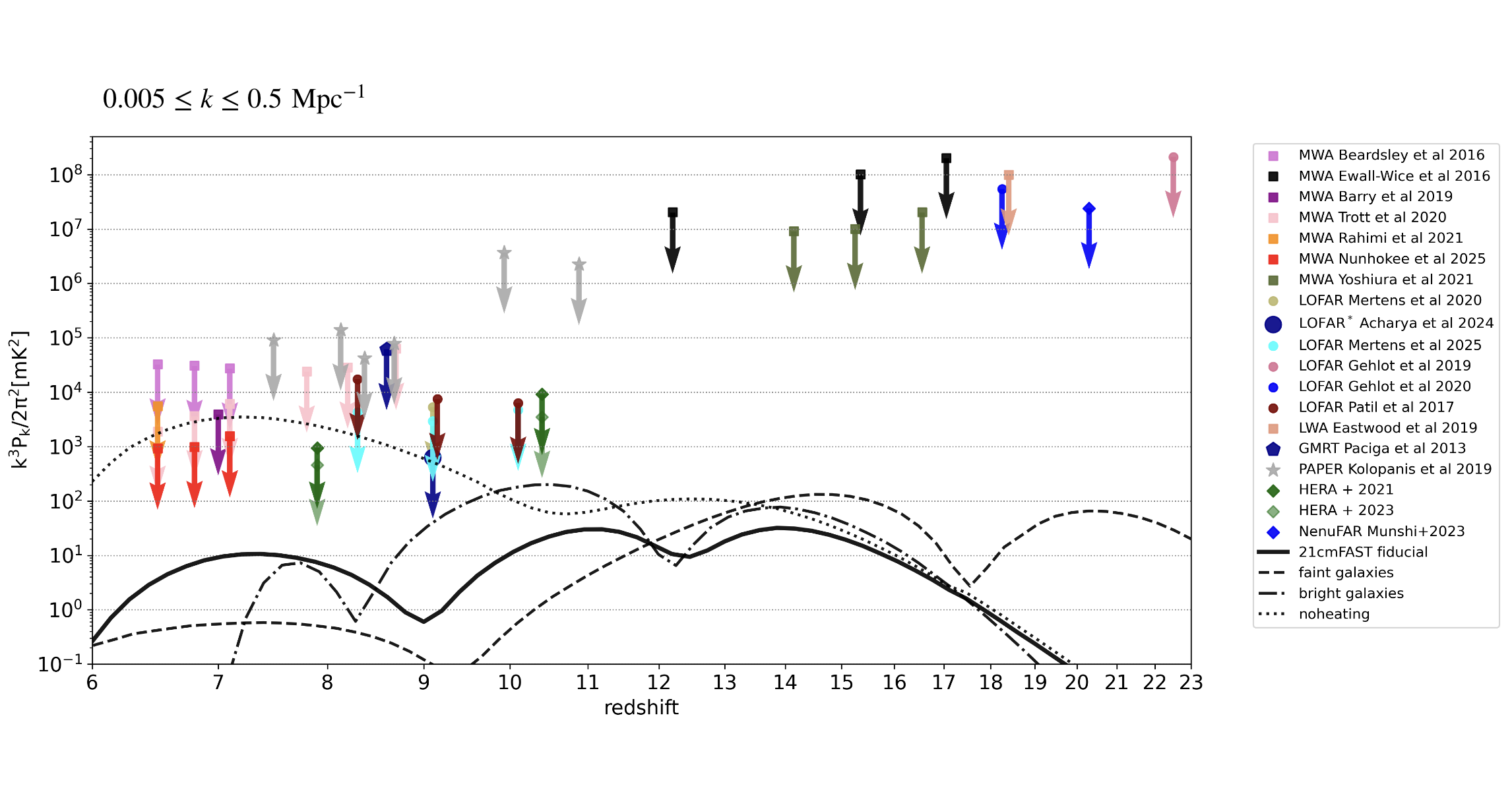}
\caption{
Comparison between theoretical predictions of the dimensionless 21 cm power spectrum, $\Delta^{2}_{21}=k^{3}P_{21}/(2\pi^{2})$, and current observational upper limits as a function of redshift over the range $0.005 \le k \le 0.5~{\rm Mpc}^{-1}$. Credit: Shintaro Yoshiura.
}
\label{fig:ps_constraints}
\end{figure}

Spatial fluctuations contain the information lost by averaging over the sky. They respond to the three-dimensional distribution of matter, radiation sources, heated regions, and ionized bubbles. A standard way to summarize these fluctuations is to Fourier transform the brightness-temperature field $\delta T_{\rm b}(\mathbf{x}, z)$ and compute its variance as a function of wavenumber. The 21 cm power spectrum is
\begin{equation}
P_{21}(k, z)
=
\left\langle
\big| \widetilde{\delta T_{\rm b}}(\mathbf{k}, z) \big|^{2}
\right\rangle,
\end{equation}
which measures the variance on different spatial scales. Observational results are often expressed using the dimensionless form
\begin{equation}
\Delta^{2}_{21}(k,z)
=
\frac{k^{3}}{2\pi^{2}} P_{21}(k,z).
\end{equation}
This is the quantity shown in Fig.~\ref{fig:ps_constraints}.

The redshift evolution of the power spectrum reflects the sequence of Lyman-$\alpha$ coupling, heating, and ionized-bubble growth. Figure~\ref{fig:ps_constraints} compares theoretical expectations for $\Delta^{2}_{21}$ with current upper limits. Current limits from MWA, LOFAR, HERA, and related pathfinder experiments remain above many fiducial predictions, but they already rule out or disfavor some extreme astrophysical scenarios and provide practical guidance for future surveys.

A power spectrum is not, however, a complete description of the 21 cm field. Reionization is highly inhomogeneous, and the resulting brightness-temperature field can be strongly non-Gaussian. For this reason, analyses often combine the power spectrum with one-point summaries, higher-order statistics, and morphology-sensitive descriptors such as skewness, bispectra, bubble statistics, or topological quantities \citep{2015MNRAS.451..467S,2015MNRAS.454.1416W,2016PASJ...68...61K,2017MNRAS.468.3785R,2018MNRAS.474.4487K,2016MNRAS.458.3003S,2017MNRAS.472.2436W,2018MNRAS.476.4007M,2020MNRAS.499.5090M,2017MNRAS.468.1542S}.

Such instruments do not measure $\delta T_{\rm b}$ directly. Instead, they measure visibilities, which correspond to Fourier components of the sky after weighting by the primary beam. Since the beam and baseline response depend on frequency, smooth-spectrum foregrounds can leak into Fourier modes that would otherwise be used for the cosmological signal. This is the reason why foreground contamination is often discussed in $(k_{\perp},k_{\parallel})$ space, leading to the familiar foreground-wedge picture\citep{2010ApJ...724..526D,2012ApJ...756..165P,2014PhRvD..90b3018L,2014PhRvD..90b3019L}.

\subsection{Beyond the power spectrum: morphology-aware summaries}

The power spectrum is a central statistic in 21 cm cosmology, but it should not be regarded as a complete description of the signal. It is useful because it is compact, closely related to interferometric measurements, and commonly used in current upper-limit analyses. However, the power spectrum is still a two-point statistic. It characterizes the scale dependence of the variance, but it does not retain the phase information or the spatial morphology of the brightness-temperature field. This is a serious limitation for the 21 cm signal, because the field is generally non-Gaussian during cosmic dawn and reionization.

This non-Gaussianity has a clear physical origin. Ionized bubbles grow around biased sources, X-ray heating can be highly inhomogeneous, and Lyman-$\alpha$ coupling also varies spatially. These processes produce a brightness-temperature field with correlated structures, such as ionized regions, heated patches, cold neutral regions, and connected neutral structures. Two maps can therefore have similar values of $P_{21}(k)$ while having different real-space morphology. In such cases, the power spectrum alone discards information that is directly relevant to the underlying astrophysics. This motivates the use of higher-order statistics and morphology-sensitive summaries in addition to $P_{21}(k)$.

A natural first step beyond the power spectrum is to study the distribution of brightness-temperature values themselves. The 21 cm probability distribution function and its moments, such as the variance, skewness, and kurtosis, summarize how the signal is distributed over a map. These statistics do not use the positions of pixels, and therefore do not describe spatial morphology. Even so, they remain sensitive to changes in the ionization and thermal state of the intergalactic medium, because ionized regions, cold neutral gas, and heated regions populate different parts of the brightness-temperature distribution. Previous studies have used the 21 cm PDF and related moments to trace reionization history and to separate models that can be partly degenerate at the level of the power spectrum \citep{2010MNRAS.406.2521I,2014MNRAS.443.3090W,2015MNRAS.451..467S,2016PASJ...68...61K}. The main merit of these statistics is that they are simple and physically interpretable.

The bispectrum gives a more direct Fourier-space probe of higher-order structure. It measures correlations among three Fourier modes and retains part of the phase-coupling information absent from the power spectrum. The 21 cm bispectrum has been used to study non-Gaussianity from X-ray heating, ionized bubbles, light-cone effects, and reionization morphology \citep{2016MNRAS.458.3003S,2018MNRAS.476.4007M,2020MNRAS.492..653H,2021MNRAS.508.3848M,2022MNRAS.510.3838W}. It is a natural statistic when the analysis stays in Fourier space, but it produces a large data vector, and its covariance and foreground response are not simple. 

Bubble-size distributions approach the problem from real space by trying to characterize the scale of ionized or neutral regions directly. This is physically well motivated, since the growth and merging of ionized bubbles is a central feature of reionization. Several definitions can be used, including spherical-average, mean-free-path, and friends-of-friends methods, and they do not always give the same answer \citep{2018MNRAS.473.2949G,2022A&A...667A.118D}. A bubble-size distribution should therefore be regarded as a convention-dependent summary rather than a unique observable. The definition of a bubble, the thresholding method, beam smoothing, and foreground filtering all affect the result.

Topological summaries are useful because they do not require choosing one bubble radius. Minkowski functionals measure geometrical and topological properties of excursion sets, such as volume, surface area, curvature, and Euler characteristic. In the 21 cm context, \citet{2017MNRAS.465..394Y} applied 3D Minkowski functionals to the brightness-temperature field and showed that they can trace non-Gaussian morphology caused by spin-temperature fluctuations and ionized bubbles. Related work using contour Minkowski tensors also studied the shape and size anisotropy of ionized regions \citep{2018JCAP...10..011K}. More recently, reionization-parameter inference with 3D Minkowski functionals has been examined with SKA-Low-like observational effects \citep{2024ApJ...974..141D}. These studies provide the most direct references for Minkowski-based 21 cm morphology.

Betti numbers and persistent homology describe topology in a different but related direction. Betti numbers count connected components, tunnels, and cavities. Persistent homology follows how these structures appear and disappear as the threshold is varied, producing a hierarchy of topological features rather than a single number at one threshold. This approach has been developed for the topology of reionization and for mock 21 cm observations \citep{2021MNRAS.505.1863G,2019MNRAS.486.1523E,2023MNRAS.520.2709E}. The main benefit is that topology describes percolation and connectivity, which are not cleanly captured by the angle-averaged power spectrum.

Wavelet-based summaries give an intermediate case between handcrafted statistics and deep neural networks. The wavelet scattering transform uses fixed wavelet filters and nonlinear operations to extract multiscale, non-Gaussian information. It is not as transparent as the power spectrum, but it is much less flexible than a trained neural network. For 21 cm images, the WST has been proposed and tested as a way to extract non-Gaussian information and improve astrophysical constraints \citep{2022MNRAS.513.1719G,2023MNRAS.519.5288G,2024A&A...686A.212H,2025arXiv251200402S}. Comparisons among different summary statistics also show that the question is not simply whether a statistic is new, but how much information it keeps after noise, foreground treatment, and parameter degeneracies are included \citep{2024A&A...688A.199P}.

These statistics are complementary because each of them compresses the 21 cm field in a different way. The power spectrum removes phase information, the PDF removes spatial information, and topology- or wavelet-based summaries retain different aspects of connectivity and multiscale structure. Therefore, agreement between different summaries is not guaranteed, and their differences can be useful for identifying which features of the non-Gaussian 21 cm field are responsible for separating astrophysical models.

The statistics discussed above can be regarded as different compressions of the same non-Gaussian 21 cm brightness-temperature field. The power spectrum remains the standard reference statistic, but it keeps only scale-dependent variance and discards the phase correlations that encode much of the morphology. One-point statistics, higher-order correlations, topological measures, and wavelet-based summaries retain different aspects of the information lost in this reduction. The purpose of these statistics is therefore not to replace $P_{21}(k)$, but to characterize the non-Gaussian structure of the 21 cm signal in ways that are not accessible from the power spectrum alone.

\subsection{21 cm forest}

The 21 cm forest provides a complementary way to study neutral hydrogen at high redshift. In contrast to tomographic 21 cm observations, which aim to measure the diffuse brightness-temperature field over the sky, the 21cm forest signal is measured as absorption features in the spectra of bright, high-redshift radio sources. These absorption features are produced by intervening neutral hydrogen along the line of sight. The observable is therefore closer to a one-dimensional transmission or optical-depth field than to a three-dimensional image of $\delta T_{\rm b}$.

This difference makes the 21 cm forest sensitive to physical scales and environments that are difficult to access with tomographic measurements alone. The absorption strength depends strongly on the spin temperature and neutral hydrogen density, and is therefore enhanced in cold neutral gas before efficient X-ray heating. It can also respond to small-scale structure, including dense absorbers and minihalo-like systems, which may leave narrow or intermittent features in individual spectra. For this reason, the 21 cm forest has been discussed as a probe of the thermal history of the neutral intergalactic medium, small-scale cosmological structure, and dark-matter microphysics \citep{2002ApJ...577...22C,2002ApJ...579....1F,2006MNRAS.370.1867F,2009ApJ...704.1396X,2011MNRAS.410.2025X,2012MNRAS.425.2988M,2014MNRAS.441.2476E,2021MNRAS.506.5818Sol,2025MNRAS.537..364Sol}.

The statistical problem is also different from that of 21 cm imaging. Early studies often focused on the abundance and depth of individual absorption lines, but the 21cm forest can also be treated as a fluctuating one-dimensional field. In this view, the relevant information is not only the number of strong lines, but also the distribution, clustering, and morphology of absorption features along the line of sight. This motivates the use of one-dimensional power spectra, halo-based modeling, and non-Gaussian summaries such as topological statistics. These approaches are useful because different thermal and structure-formation models can produce spectra with similar average absorption but different small-scale morphology \citep{2014PhRvD..90h3003S,2020PhRvD.101d3516S,2020PhRvD.102b3522S,2023PhRvD.107l3520S,2023NatAs...7.1116S,2025PhRvD.112f3513S,2025PhRvD.112f3557S,2026PhRvD.113h3525S,2025MNRAS.537..364Sol}.

The main limitation is observational. Detecting the 21 cm forest requires sufficiently bright radio-loud sources at high redshift and enough spectral sensitivity to resolve weak absorption features. The method is therefore not a guaranteed alternative to tomographic 21 cm observations. Rather, it is a complementary probe. If suitable background sources are available, the 21cm forest can access small-scale and cold-gas information.

\section{Challenges in interpreting 21 cm data}

Interpreting 21 cm data is difficult because several problems occur simultaneously: the signal is intrinsically non-Gaussian, the measurements are contaminated by severe foregrounds and instrumental effects, and inference requires expensive forward modeling.

\subsection{Physical and computational challenges}

The brightness temperature depends on several coupled fields: density, ionization fraction, kinetic and spin temperatures, and radiation backgrounds. These fields evolve nonlinearly and are driven by source populations whose properties are still poorly constrained. As a result, different astrophysical models can produce similar global signals or power spectra \citep{2011MNRAS.411..955M,2015MNRAS.449.4246G,2017MNRAS.472.2651G,2018MNRAS.477.3217G,2019MNRAS.484..933P,2014MNRAS.438.2664S,2014MNRAS.439.3262M,2023NatAs...7.1116S,2025PhRvD.112f3557S}.

These degeneracies are not only degeneracies between parameter values. Information is distributed across redshift, scale, and morphology. Two models may agree in the global signal or in the power spectrum while differing in ionized-bubble topology, higher-order statistics, or light-cone anisotropy. This is one reason to use multiple summary statistics, or field-level information, rather than compressing the signal too early.

Computation creates a second bottleneck. Robust inference requires many forward simulations and a reliable treatment of uncertainties in large data vectors. Semi-numerical simulations are much faster than full radiation-hydrodynamic calculations, but Bayesian exploration can still become expensive when repeated over large astrophysical parameter spaces. The difficulty increases when realistic covariance, foreground choices, instrumental effects, and analysis cuts are included \citep{2024A&A...688A.199P}.

These complications affect not only the computational cost but also the reliability of the final constraints. Calibration residuals, foreground modeling, beam uncertainties, masking, and data-selection choices can change the recovered signal. Similarly, covariance matrices estimated from a finite number of simulations are noisy, and semi-numerical models do not reproduce all features of higher-fidelity calculations. If these effects are treated as fixed or negligible, the resulting constraints can appear artificially precise. Emulator errors, simulator mismatch, and model discrepancy should therefore be included in the uncertainty budget, or at least tested through robustness checks \citep{2025MNRAS.544..375B}.

\subsection{Observational complexity}

Foreground emission is brighter than the cosmological 21 cm signal by many orders of magnitude. Its intrinsic spectral smoothness provides an important handle for separation, but the response of a low-frequency interferometer is itself chromatic. The primary beam and projected baselines vary with frequency, so foreground emission that is smooth on the sky can acquire spectral structure in the measured visibilities. This mode mixing is the origin of foreground leakage into higher $k_{\parallel}$ modes and of the foreground wedge \citep{2019arXiv190912369C,2010ApJ...724..526D,2012ApJ...756..165P,2014PhRvD..90b3018L,2014PhRvD..90b3019L,2020PASP..132f2001L}.

The visibility domain is therefore the natural place to understand many observational systematics. An interferometer does not measure the brightness-temperature field directly, but samples Fourier components of the beam-weighted sky. This makes the measurement sensitive to incomplete $uv$ coverage, chromatic sidelobes, direction-dependent gains, and beam-model errors. Radio-frequency interference, ionospheric variations, gain-calibration errors, polarization leakage, and incomplete sampling add further complications \citep{2010MNRAS.405..155O,2015PASA...32....8O,2016RaSc...51..927M,2018A&A...615A.179D,2016PASA...33...19T,2020MNRAS.493.1662M,2021MNRAS.501.1463K,2024MNRAS.533..632B,2024MNRAS.527.3517M}.

Calibration errors are especially problematic when they introduce chromatic residuals. Small gain errors, beam inaccuracies, direction-dependent effects, or polarization leakage can generate spectral structure that overlaps with the modes used for the cosmological signal. RFI flagging creates a related problem by making the sampling in time and frequency irregular. If the resulting window functions and error propagation are not modeled carefully, the irregular sampling can broaden mode mixing and complicate the interpretation of the recovered power spectrum.

Foreground mitigation therefore involves a trade-off. Wedge avoidance is conservative because it discards regions of Fourier space that are expected to be foreground contaminated, but it also removes cosmological modes. Foreground subtraction or reconstruction can recover more information, but it depends more strongly on sky modeling, beam modeling, and calibration accuracy. For SKA-Low analyses, a combination of foreground avoidance, subtraction or reconstruction, and validation at the visibility level will likely be needed rather than reliance on a single foreground strategy.

\subsection{Additional SKA-era challenges}

SKA-Low will change the scale of 21 cm analysis. The data volume, calibration complexity, imaging choices, foreground treatment, summary-statistic estimation, and parameter inference will all be connected through a long processing chain. Choices made at an early stage can affect the data vector and its uncertainty at later stages. For this reason, pipeline architecture is not only an engineering issue, but also part of the scientific analysis \citep{2021MNRAS.504.4716G,2024MNRAS.529.3684K,2024arXiv241215893S}.

At this scale, reproducibility becomes a central requirement. Continuous visibility data, large calibration state spaces, repeated imaging, and multiple foreground-mitigation choices make manually tuned offline analyses difficult to control. A reliable analysis pipeline needs provenance tracking, automated quality assessment, deterministic reprocessing, and a clear record of how calibration, flagging, imaging, and foreground choices propagate into the final data products.

These requirements are especially important because future 21 cm constraints will depend on small residual signals extracted after many processing steps. It will not be enough to report a final power spectrum or parameter constraint. The analysis must also show how the result changes under reasonable variations of calibration models, foreground treatment, data selection, and error propagation. In this sense, SKA-era 21 cm cosmology requires not only more sensitive data, but also analysis workflows that are reproducible, scalable, and able to carry uncertainties through the full pipeline.



\section{Machine learning approaches to 21 cm signal analysis}

Machine-learning(ML) methods are now used in many parts of 21 cm cosmology \citep{2023RPPh...86g6901M}. They should not be treated as a single class of tools. A method used for RFI flagging, a Gaussian-process model for foreground covariance, an emulator for simulated power spectra, and a neural posterior estimator for astrophysical parameters address different problems. They also require different training data, validation tests, and performance metrics.

\begin{figure}[htbp]
\centering
\includegraphics[width=0.95\linewidth]{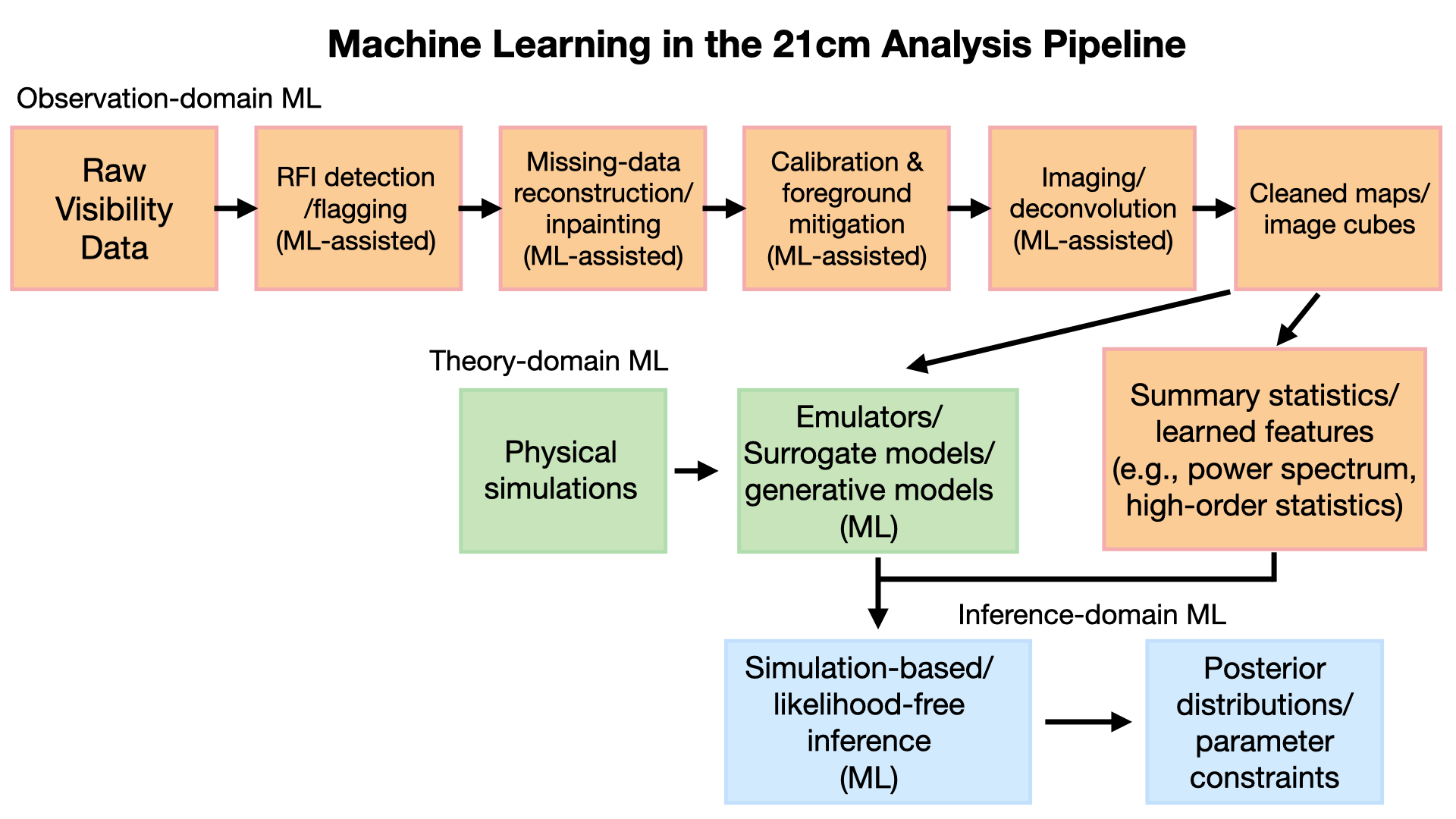}
\caption{
Schematic overview of ML workflows in 21 cm cosmology.
The arrows indicate the main forward flow emphasized in this review: ML can enter at the observation stage to clean or reconstruct contaminated data products, at the theory stage to emulate or compress forward models, and at the inference stage to map observables to astrophysical or cosmological constraints.
In practice, validation and iterative refinement can also pass information in the reverse direction.
}
\label{fig:ml_pipeline_overview}
\end{figure}

For this reason, this section organizes the literature by the role that each method plays in the analysis pipeline. Some methods act on observational data, for example by flagging, calibration support, foreground modeling, or residual identification. Others are used on the theory side, where they emulate simulations or compress high-dimensional summary statistics. A third class is used for inference, where data products are mapped to astrophysical or cosmological parameters. The boundaries between these categories are not strict, but this organization makes clear what each ML method is expected to do and how it should be tested. Figure~\ref{fig:ml_pipeline_overview} summarizes this workflow.

\subsection{Observation-domain methods}

Observation-domain ML refers to methods applied before the final cosmological inference. The input may be visibilities, time--frequency data, dirty maps, image cubes, or intermediate data products, and the output is usually a cleaned, reconstructed, flagged, or compressed data product rather than a posterior distribution. This distinction is important because the effect of such methods must be evaluated through the later analysis steps. A map that is visually cleaner, or a cube with smaller apparent variance, is useful only if the 21 cm statistics used for science are preserved.

RFI mitigation is one of the main applications at this stage. Supervised classifiers and anomaly-detection methods have been used to flag contaminated time--frequency samples more systematically than manual inspection or simple thresholding \citep{2017A&C....18...35A,2019MNRAS.488.2605K,2020MNRAS.499..379V}. The relevant output, however, is not only the set of flagged samples. Flagging changes the sampling window that enters map-making and power-spectrum estimation. False positives remove usable data, while false negatives leave residual contamination. Both effects can propagate into mode mixing, noise estimates, and signal loss.

The treatment of missing samples is therefore closely related to RFI mitigation. Hard flagging leaves gaps in time--frequency space, and these gaps can broaden Fourier leakage if the corresponding window function is not modeled accurately. Learned inpainting or reconstruction has been proposed as a way to fill missing regions before subsequent analysis \citep{2023MNRAS.520.5552P}. For this use case, the method should be viewed as an interpolation procedure constrained by the statistics of the data. Its performance should be tested not only by reconstruction error in the missing pixels, but also by the effect on the power spectrum, covariance, and any non-Gaussian summaries used later.

Foreground mitigation is another major observation-domain application. Because the chromatic instrument response mixes foreground power into cosmological modes, the task is not simply to remove a spectrally smooth component. The method must account for the joint angular and spectral structure of the foreground residuals while limiting the loss of the 21 cm signal. Neural networks have been used to reconstruct wedge-filtered light-cones and to recover information from contaminated tomographic data products \citep{2021MNRAS.504.4716G,2024MNRAS.529.3684K}. Gaussian-process foreground pipelines and ML-assisted covariance models have also been developed for more flexible component separation, including applications to LOFAR EoR analyses \citep{2024MNRAS.527.7835A,2024MNRAS.527.3517M,2024MNRASL.534L..30A}. In these applications, the key quantities to monitor are the residual foreground covariance and the amount of cosmological signal removed by the cleaning procedure.

Calibration and propagation effects provide a related class of problems. Gain errors, beam mismatch, direction-dependent effects, polarization leakage, and ionospheric variations can introduce chromatic residuals that overlap with the modes used for the 21 cm signal. These residuals are structured, but they are often difficult to describe with a small number of analytic parameters. Data-driven corrections and adaptive covariance models have therefore been explored as ways to reduce leakage and stabilize recovered power spectra \citep{2021MNRAS.501.1463K,2024MNRAS.533..632B,2024MNRAS.527.3517M}. For global-signal radiometers, the measurement is not interferometric, but the role of calibration is similar. Receiver reflections and impedance mismatch can introduce spectral structure in a sky-averaged measurement, and ML-based calibration frameworks have been tested for such systems \citep{2025SciRep..1534335L}.

Image-domain methods address a different type of data product. Segmentation networks have been used to identify ionized and neutral regions in noisy or foreground-degraded maps, enabling measurements based on bubble sizes, morphology, or topology \citep{2021MNRAS.505.3982B,2024MNRAS.528.5212B}. Related work has considered partial recovery of the 21 cm morphology from foreground-contaminated observations \citep{2025MNRAS.541..234B}. Since reionization produces a non-Gaussian brightness-temperature field, image-domain methods can in principle retain information that is not captured by the power spectrum alone. Their validation should therefore be tied to the intended scientific summary: for example, bubble-size distributions, Minkowski functionals, Betti curves, wavelet coefficients, or downstream parameter constraints, rather than only pixel-level agreement.

Observation-domain ML can also be used with external tracers. Conditional generative models have been applied to predict 21 cm maps from Lyman-$\alpha$ emitter distributions, and related image-to-image methods have been used in line-intensity mapping to separate confused components and reconstruct three-dimensional fields \citep{2021MNRAS.506..357Y,2020MNRAS.496L..54M,2021ApJ...906L...1M,2021ApJ...923L...7M}. These methods are best described as cross-probe conditioning or reconstruction methods. They can provide auxiliary information about the expected morphology of the 21 cm field, but they do not replace the radio measurement itself.

Across these examples, the validation criterion depends on where the ML component enters the analysis chain. For RFI flagging, the relevant test is the induced sampling window and its effect on leakage. For inpainting and foreground reconstruction, it is the preservation of the statistics used in later analysis. For calibration-related applications, it is the reduction of chromatic residuals without removing cosmological signal. For image-domain reconstruction, it is the recovery of the morphology or summary statistics used for interpretation. Observation-domain ML should therefore be evaluated through the data products and summary statistics that enter the final 21 cm analysis.

\subsection{Theory-domain methods}

On the theory side, ML is mainly useful for repeated forward modeling. The relevant question is not whether a network recognizes a pattern, but which expensive calculation can be replaced or accelerated without losing the physics needed for inference. In 21 cm cosmology, the most established answer is summary-statistic emulation.

An emulator replaces many calls to a forward model with a fast surrogate trained on a suite of simulations. Early studies used neural networks and Gaussian processes to interpolate the 21 cm power spectrum across astrophysical parameter space, allowing Bayesian exploration to proceed much faster than brute-force simulation \citep{2017ApJ...848...23K,2018MNRAS.475.1213S,2019MNRAS.483.2907J}. This remains one of the clearest uses of ML in the field. Its strength is speed, but an emulator of a weak summary cannot recover information that the summary has already discarded.

Global-signal emulation has become a distinct subfield. The 21cmGEM emulator of \citet{2020MNRAS.495.4845C} was an early neural-network emulator of the global signal from cosmic dawn and reionization. The later \texttt{globalemu} framework used redshift as an input variable and provided a fast emulator for the sky-averaged signal and neutral-fraction history \citep{2021MNRAS.508.2923B}. Related work then explored different neural representations, including 21cmVAE, recurrent long short-term memory networks, and Kolmogorov--Arnold networks \citep{2022ApJ...930...79B,2024ApJ...977...19D,2025ApJ...991..152D}. The global signal is a one-dimensional curve, but its frequency channels are not independent data points. They trace an ordered sequence set by the thermal and ionization history, and an emulator that accounts for this ordering can retain information that may be lost when the channels are treated separately.

Emulation is also no longer limited to one observable. \citet{2024MNRAS.527.9833B} developed 21cmEMU to emulate several 21cmFAST summary observables, including the neutral fraction, power spectrum, spin temperature, global signal, UV luminosity functions, and CMB optical depth. Such tools matter because realistic constraints rarely come from one measurement alone. They also connect naturally to analyses of HERA power-spectrum limits, where forward modeling and statistical validation already shape the interpretation of upper limits \citep{2023ApJ...945..124H}.

A deeper issue is compression. The power spectrum is widely used because it is well defined, relatively low dimensional, and directly connected to current upper-limit analyses. However, when the 21 cm field becomes non-Gaussian, the choice of summary becomes part of the scientific model. One response is to enrich the handcrafted summary set, for example, by combining multiple redshifts, higher-order statistics, or morphology-sensitive quantities. Another response is to learn compact representations from simulated maps or light-cones, and then to test whether they improve parameter recovery. Recent studies comparing multiple summaries within a common inference framework suggest that neither traditional nor learned compression should be preferred in advance. The useful criterion is retained parameter information at fixed computational cost \citep{2024A&A...688A.199P,2025A&A...698A..35S}.

Light-cone and map-based neural studies are therefore important even when they are presented as inference papers. They show that morphology, redshift ordering, and non-Gaussian texture carry information beyond low-dimensional summaries \citep{2019MNRAS.484..282G,2022MNRAS.509.3852P,2022MNRAS.511.3446N,2022ApJ...926..151Z,2022ApJ...933..236Z}. They also indicate what theorists may need to emulate next. The target may not be only the power spectrum, but a representation that preserves the physically relevant structure of the field.

Fully field-level surrogates and generative field models remain less developed than power-spectrum emulators. Interpolating a low-dimensional observable is much easier than generating three-dimensional fields with correct topology, redshift evolution, and parameter dependence. Cross-domain image-to-image studies and related generative models indicate possible directions, but they are not yet substitutes for physical simulators \citep{2021MNRAS.506..357Y,2020MNRAS.496L..54M,2021ApJ...906L...1M,2021ApJ...923L...7M}. For these models, visual realism is a weak criterion. A field-level surrogate should be judged by whether it preserves non-Gaussian observables, topology, redshift evolution, and parameter dependence when the input moves away from the densest part of the training set.

Another promising, but still less mature, route is operator-level acceleration. Instead of replacing the whole simulator, a costly substep is emulated, or a residual correction around a physical approximation is learned. This is useful because it localizes the approximation and can make error accounting easier than in one monolithic generator. However, in 21 cm cosmology, this approach is still not as developed as whole-observable emulation. It should not be hidden under the broad phrase ``ML surrogate modeling'' without specifying what is actually emulated.

Emulators are useful only when their errors are characterized together with their predictions. In parameter inference, an inaccurate surrogate can change the likelihood and shift the recovered posterior, even if its average prediction error appears small. Emulator uncertainty should therefore be propagated into the final constraints, or represented through an explicit discrepancy term between the emulator and the underlying simulator. Recent work has quantified how emulator inaccuracies affect posterior recovery, rather than evaluating emulator performance only with a root-mean-square prediction error \citep{2025MNRAS.544..375B}. In theory-domain applications, the emulator should be treated as an approximation to a specified forward model, with its residual error included in the uncertainty budget.

\subsection{Inference-domain methods}

Inference-domain ML uses simulated observables to infer astrophysical or cosmological parameters. Some methods give deterministic outputs, such as class labels or point estimates of parameters. Others approximate a posterior distribution, a likelihood, or a likelihood ratio. The distinction matters because these approaches answer different statistical questions and require different validation tests.

Early applications of ML to 21 cm cosmology mainly used classifiers and regressors. These studies demonstrated that maps and light-cones contain information beyond conventional two-point summaries. For example, CNN classifiers were used to distinguish AGN-driven and galaxy-driven reionization from morphology alone \citep{2019MNRAS.483.2524H}. Regression models were then applied to recover astrophysical parameters from power spectra, images, and light-cones, often with good accuracy in idealized tests \citep{2017MNRAS.468.3869S,2019MNRAS.484..282G,2022MNRAS.512.5010C}. Later studies extended this approach to contaminated light-cones, alternative network architectures, and joint inference of astrophysical parameters with dark-matter physics \citep{2022MNRAS.509.3852P,2022MNRAS.511.3446N,2022ApJ...926..151Z,2022ApJ...933..236Z,2023MNRAS.525.6097S}. Related artificial neural network studies also showed that derived quantities, such as bubble-size statistics and the global 21 cm signal, can be reconstructed from other summaries \citep{Shimabukuro2022RAA,Shimabukuro2025RAA}.

Global-signal inference has a different structure. The data vector is low dimensional, but foreground and instrumental terms dominate the analysis \citep{2019arXiv190912369C,2020PASP..132f2001L}. ML has therefore been used as a flexible emulator for the physical signal, as a model for nuisance components, and as an amortized tool for joint inference over astrophysical and instrumental parameters \citep{2020MNRAS.495.4845C,2021MNRAS.508.2923B,2022ApJ...930...79B,2024ApJ...977...19D,2025ApJ...991..152D}. In this application, validation has to include the foreground model, beam chromaticity, calibration drift, and other instrumental effects, not only the recovery of the astrophysical signal in idealized simulations.

Deterministic regression remains useful for testing what information is present in a data vector and for constructing compressed representations. A point estimate, however, does not specify a calibrated uncertainty. This has led to increasing use of simulation-based inference and related likelihood-free methods in 21 cm analysis \citep{2024A&A...688A.199P,2025A&A...698A..35S,2019MNRAS.488.4440A,2020PNAS..11730055C}. Depending on the method, the neural model approximates the posterior, the likelihood, or a likelihood ratio from simulations, and can then be applied repeatedly to mock or observed data vectors.

\begin{figure}[htbp]
\centering
\includegraphics[width=0.92\linewidth]{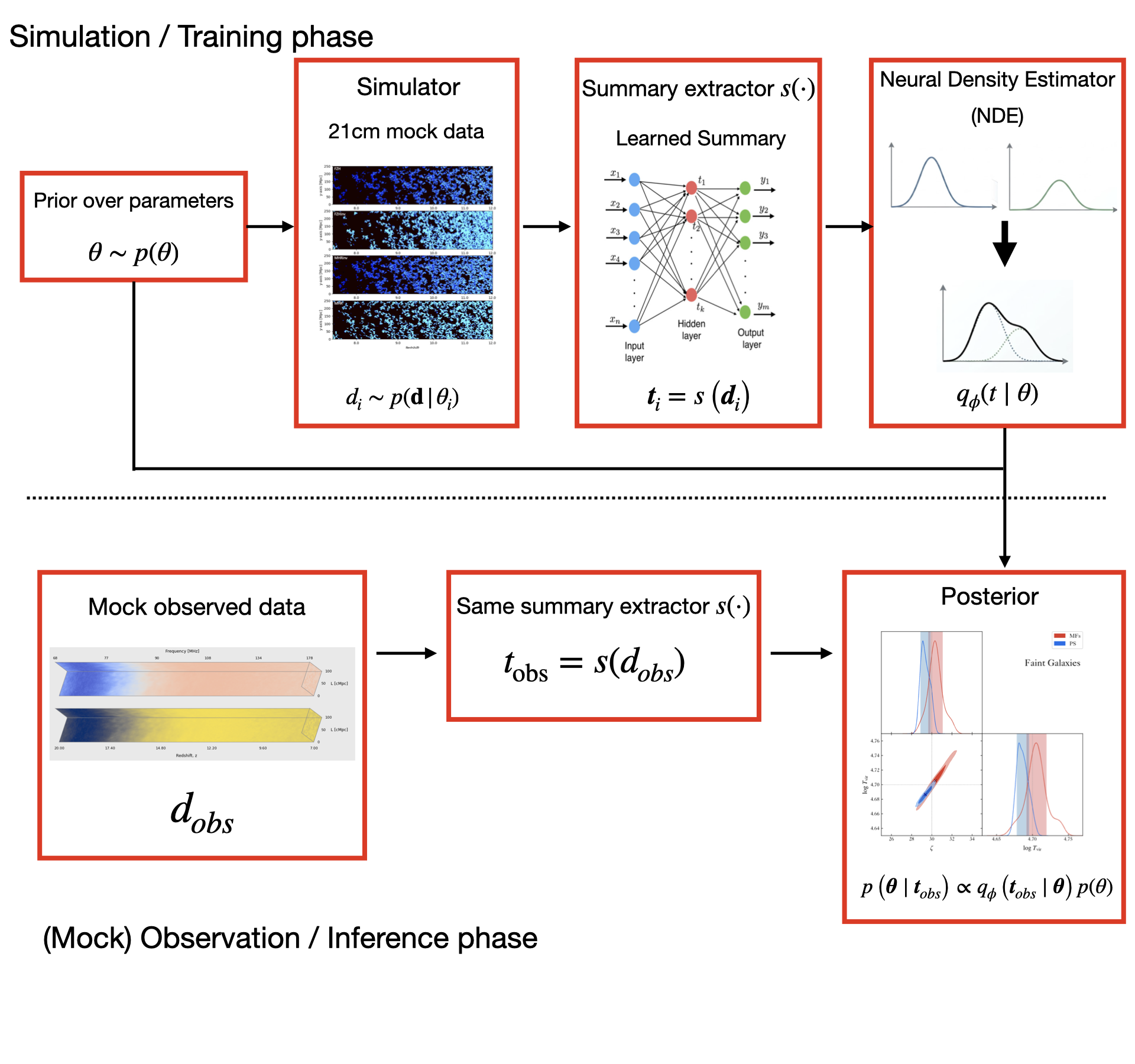}
\caption{
Illustrative schematic of ML-based likelihood-free parameter inference for 21 cm cosmology. Astrophysical and cosmological parameters are mapped through forward modeling to observables such as the global signal, power spectrum, maps, or light-cones, and neural inference modules then learn posterior constraints directly from these data products while accounting for uncertainty and model complexity. Parts of this figure are reproduced from \citep{2024ApJ...974..141D,2017MNRAS.468.3869S}.
}
\label{fig:lfi_21cm_workflow}
\end{figure}

A typical simulation-based inference(SBI) workflow draws parameters from a prior, runs the simulator, computes observables, and trains a conditional density estimator using the resulting parameter--data pairs. Neural posterior estimation, neural likelihood estimation, and neural ratio estimation differ in the quantity that they approximate. In practice, they often use flexible density models such as normalizing flows or autoregressive networks \citep{2015arXiv150505770R,2017arXiv170507057P}. Figure~\ref{fig:lfi_21cm_workflow} shows this logic schematically for 21 cm cosmology.

Neural density estimators are also useful when heterogeneous data sets are combined. \citet{2024MNRAS.527..813B} combined SARAS3 global-signal information with HERA power-spectrum limits. More recent work used neural-density-estimation-accelerated Bayesian analysis to combine 21 cm power-spectrum upper limits with non-21 cm probes, including CMB and Lyman-line constraints \citep{2025MNRAS.544.3856S}. Such analyses are relevant for future constraints because the first-galaxy parameter space will likely be constrained by several probes and summary statistics rather than by a single measurement.

Within SBI, the data representation does not have to be a single summary statistic. The inference model can condition on power spectra, multiple redshift slices, learned embeddings of maps or light-cones, or combinations of physically motivated and learned summaries. Recent work has shown that combining summaries within an SBI framework can tighten constraints relative to using one summary alone \citep{2025A&A...698A..35S}. This reflects the structure of the 21 cm signal, where information is distributed across scale, redshift, and morphology.

Recent comparison studies have examined how the inferred constraints depend on the simulator, the data representation, and the probabilistic estimator. A flexible density estimator can only use the information present in the summary supplied to it, and its posterior will still reflect any bias in the simulator used for training. Comparisons based on common simulation databases are therefore useful for testing calibration, coverage, and robustness under shared conditions, rather than only for ranking network architectures \citep{2025A&A...698A..80M,2025A&A...698A..35S}.

Posterior calibration is a central issue for inference-domain ML. SBI can produce smooth posteriors even when the simulator, emulator, or noise model is incomplete. Recent SKA-oriented work has therefore studied simulator mismatch, domain shift, coverage failure, and emulator-induced posterior bias in the context of 21 cm inference \citep{2024arXiv241215893S,2025MNRAS.544..375B}. Useful diagnostics include coverage tests, simulation-based calibration, posterior predictive checks, and tests with alternative simulators \citep{2018arXiv180406788T}. Transfer-learning studies between idealized simulations and more realistic mock observations also show that performance can degrade when the training simulations and the target data differ in foreground treatment, noise, or instrumental response \citep{2022MNRAS.511.3446N}. Performance on held-out simulations is therefore only one part of the validation; the posterior should also be tested against changes in the simulator, data representation, and observational assumptions.

\subsection{Machine learning for the 21 cm forest}

The 21 cm forest has a different data structure from tomographic 21 cm imaging. It is observed as absorption structure in spectra of bright high-redshift radio sources, rather than as a three-dimensional diffuse brightness-temperature map. The relevant information is encoded along one-dimensional sightlines, through the depth, width, distribution, and correlation of absorption features. For this reason, methods developed for two- or three-dimensional 21 cm images cannot be transferred to forest spectra without changing the data representation.

Machine-learning applications to the 21 cm forest are still more limited than those for 21 cm imaging or global-signal analyses. Recent studies have focused mainly on learning compressed representations of noisy spectra and using them for parameter inference. In this setting, the input is not an image cube but a set of line-of-sight spectra, and the aim is to retain information that may be weak or distributed over many absorption features. Learned summaries and latent encodings have been shown to improve the recovery of information about thermal history and dark-matter physics compared with analyses based only on a simple one-dimensional power spectrum, especially in low-SNR regimes \citep{2025CmPhy...8..220Sun,2026MNRAS.546.stag236P}.

This application is physically motivated because the 21 cm forest is sensitive to small-scale structure and to the thermal state of neutral gas. Warm dark matter, small-scale matter suppression, gas temperature, X-ray heating, and the abundance of compact neutral structures can all affect the absorption pattern \citep{2014PhRvD..90h3003S,2020PhRvD.101d3516S,2020PhRvD.102b3522S,2023PhRvD.107l3520S,2025PhRvD.112f3557S}. The same absorption spectrum can therefore be affected by several physical effects at once. This is one reason why recent forest analyses have moved beyond counting individual absorption lines or using only a one-dimensional power spectrum.

Closely related to these ML-based studies, topological summaries have also been applied to 21 cm forest spectra. Persistent homology describes how absorption troughs appear and merge as the threshold is varied, giving a compact description of the organization of absorption systems along frequency or comoving distance. This is not a machine-learning method by itself, but it provides an interpretable representation that can be used together with regression or simulation-based inference. Recent work indicates that such topological summaries can provide information complementary to amplitude-based statistics when thermal history and dark-matter effects are partially degenerate \citep{2026PhRvD.113h3525S}.

The usefulness of these methods also depends on the availability of suitable background sources. The number density, spectra, variability, and redshift distribution of sufficiently bright high-redshift radio sources remain uncertain \citep{2021MNRAS.506.5818Sol,2025MNRAS.537..364Sol}. This uncertainty is not specific to ML, but it affects how forest-based inference should be interpreted. A method tested on simulated spectra assumes a source population, a continuum model, and an observing setup, and these assumptions enter the inferred sensitivity to thermal history or dark-matter physics.

Thus, the existing ML literature on the 21 cm forest is mainly concerned with representation learning and inference from simulated sightline spectra. Its role is to extract information from weak and non-Gaussian absorption structure that is not fully captured by a one-dimensional power spectrum. At present, this is a narrower and more specialized literature than ML applications to 21 cm images, foreground mitigation, or global-signal inference.

\section{Summary}

This chapter reviewed ML applications in 21 cm cosmology by focusing on where each method enters the analysis pipeline. Observation-domain methods are closest to direct operational use for tasks such as RFI mitigation, foreground and covariance modeling, calibration support, and morphology-aware reconstruction. Theory-domain methods are strongest when they accelerate a specified forward model, especially by emulating summary statistics. Inference-domain methods range from fast deterministic regressors to SBI frameworks that aim to return calibrated posteriors for non-Gaussian observables.

The main message is that the 21 cm signal should not be compressed too early into a single statistic. The power spectrum remains a central observable, but cosmic-dawn and reionization fields contain non-Gaussian and morphological information. PDFs, higher-order statistics, topological summaries, wavelet-based summaries, and learned representations are useful because they probe different aspects of the same field. The 21 cm forest makes the same point in another form: its spectra are one-dimensional, but they carry small-scale information and are strongly affected by the uncertain population of background radio sources.

The evidence is strongest for targeted ML components with defined inputs, outputs, and validation tests. It is weaker for broad claims that end-to-end networks or generative models can replace large parts of the physical analysis pipeline. For SKA-Low, the most defensible strategy is to use ML where it makes a specific subproblem faster or better conditioned, while keeping the physical model, uncertainty budget, and validation tests visible. A learned compression or reconstruction is valuable only if it preserves the information needed for inference and remains calibrated under realistic changes in foregrounds, instruments, and simulations.



\end{document}